\documentclass[twocolumn,showpacs,amsmath,amssymb]{revtex4}
\usepackage{natbib}
\usepackage{graphicx}
\usepackage{dcolumn}
\usepackage{bm}
\newcommand{\bra}[1]{\langle#1|}
\newcommand{\ave}[1]{\langle#1\rangle}
\newcommand{\modul}[1]{|#1|}
\newcommand{\ket}[1]{|#1\rangle}

\begin{document}

\title{NONLOCAL PROPERTIES OF ENTANGLED TWO-PHOTON GENERALIZED
BINOMIAL STATES IN TWO SEPARATE CAVITIES}

\author{R. Lo Franco, G. Compagno, A. Messina, and A. Napoli}
\affiliation{Dipartimento di Scienze Fisiche ed Astronomiche,
Universit\`a di Palermo, via Archirafi 36, 90123 Palermo, Italy\\
email: lofranco@fisica.unipa.it}

\date{\today}
\pacs{42.50.Dv, 03.65.Ud, 03.67.Mn}

\begin{abstract}
\textbf{Abstract}---We consider entangled two-photon generalized
binomial states of the electromagnetic field in two separate
cavities. The nonlocal properties of this entangled field state are
analyzed by studying the electric field correlations between the two
cavities. A Bell's inequality violation is obtained using an
appropriate dichotomic cavity operator, that is in principle
measurable.
\end{abstract}

\maketitle

\section{Introduction}
Entanglement of spatially separate quantum systems holds an
important role both in the investigation of quantum theory
foundations \cite{epr} and in quantum information and computation
processing \cite{niels}. Manifestations of quantum nonlocal
properties of entangled distant systems can be characterized by
Bell's inequality violations \cite{bell,clau}. Therefore, for their
striking quantum nonlocal properties and their possible
applications, entangled quantum systems are subject of intense study
in several contexts.

In particular, in cavity quantum electrodynamics (CQED) several
schemes have been proposed to generate entangled number states in
two separate single-mode high-$Q$ cavities, the cavities having zero
or one photon \cite{rai,bergou,mey,brow,mes,compagno1}. There are
some proposals to prove Bell's inequality violations for such states
\cite{gerry,kim} but an experimental test has not yet been made. It
appears also of interest to obtain entanglement between
electromagnetic field states with mesoscopic characteristics, so
that the classical-quantum border may be approached. This is
possible, for example, if the electromagnetic field states present
non-zero mean fields. This fact seems to exclude the number states,
and some schemes to generate entangled coherent states in two
separate cavities have been then proposed \cite{davi,har3}. However,
since two different coherent states are never orthogonal, entangled
states of this kind cannot be made totally distinguishable.
Therefore, it may be useful to have entangled states in separate
cavities formed by field states that present non-zero mean fields,
are orthogonal and can be obtained by standard resonant interactions
of two-level atoms with the cavities.

Nonclassical states of electromagnetic field suited for this goal
are, for example, the generalized binomial states
\cite{sto,mou,lof2}. These states, characterized by a finite maximum
number of photons $N$, interpolate between the coherent state and
the number state and present non-zero mean electric fields.
Moreover, for each $N$-photon generalized binomial state ($N$GBS) it
is always possible to find an orthogonal one \cite{lof}. For large
$N$, these states thus present mesoscopic properties. The point
remains on how to generate entangled orthogonal couples of these
binomial states and characterize their nonlocal properties. It has
been recently indicated how both entangled 1GBSs and 2GBSs
\cite{lof,lof1} can be generated in two separate cavities by
resonant atom-cavity interactions. In the one-photon case, the
electric field correlations of the two cavities have been analyzed
and Bell's inequality violations are shown to be observable
\cite{lof}.

Since entangled 2GBSs may be generated, it appears of interest to
analyze their nonlocal properties and compare them with the ones of
the one-photon case. This constitutes the aim of this paper. In
particular, we shall examine the correlations of electric field for
entangled 2GBSs and, by introducing an appropriate dichotomic cavity
operator, we also construct a Bell's inequality that it is shown to
be violated for a wide range of the degree of entanglement and is
amenable to an experimental verification.

The paper is organized as follows. In Sec.~\ref{GBS} we recall the
definition of generalized binomial state and some of their useful
properties. In Sec.~\ref{Ent2GBS} we define the entangled two-photon
generalized binomial states, briefly describing their possible
generation scheme \cite{lof1}. In Sec.~\ref{elcorr} we study the
electric field correlations, while in Sec.~\ref{bellviol} we show
Bell's inequality violations for these entangled states. In
Sec.~\ref{concl} we summarize our conclusions.

\section{Generalized binomial state\label{GBS}}
The normalized $N$-photon generalized binomial state ($N$GBS) is
defined as \cite{sto}
\begin{equation}
\ket{N,p,\phi}\equiv\sum_{n=0}^N\left[{N\choose
n}p^{n}(1-p)^{N-n}\right]^{1/2}e^{in\phi}\ket{n},\label{bin}
\end{equation}
where $0\leq p\leq1$ is the probability of single photon occurrence,
$\phi$ the mean phase \cite{vid} and ${N\choose n}=N!/[(N-n)!n!]$.
As said above, the $N$GBS of Eq.~(\ref{bin}) interpolates between
the number state and the coherent state. In fact, for $p=0,1$ it is
reduced to the number states $\ket{0}$, $\ket{N}$ respectively. On
the other hand, when $N\rightarrow\infty$ and $p\rightarrow0$,
fixing $Np\equiv|\alpha|^2$, the $N$GBS reduces to the coherent
state $\ket{\alpha}$ with $\alpha=|\alpha|e^{i\phi}$. Two $N$GBSs
$\ket{N,p,\phi}$, $\ket{N,p',\phi'}$ are orthogonal if and only if
\cite{lof}
\begin{equation}
p'=1-p,\quad\phi'=(2k+1)\pi+\phi\quad (\textrm{$k$
integer})\label{ortog}.
\end{equation}

\section{Entangled two-photon generalized binomial states\label{Ent2GBS}}
Since it is possible to generate, by opportune resonant interactions
between two-level atoms and cavities, entangled 2GBSs in two
separate cavities \cite{lof1}, we assume here that two identical
separate single-mode cavities, namely $C_1,C_2$, are prepared in the
state
\begin{eqnarray}
\ket{\Psi^{(2)}}&=&\mathcal{N}\big[\ket{2,p_1,\phi_1}_1\ket{2,1-p_2,\pi+\phi_2}_2\nonumber\\
&+&\eta\ket{2,1-p_1,\pi+\phi_1}_1\ket{2,p_2,\phi_2}_2\big],\label{entbin2}
\end{eqnarray}
where $\eta$ is taken real, ${\cal N}=1/\sqrt{1+|\eta|^2}$ is a
normalization constant and $\ket{2,p_j,\phi_j}_j$ ($j=1,2$) is the
2GBS inside $C_j$, as obtained by Eq.~(\ref{bin}). Since each couple
of 2GBSs of Eq.~(\ref{entbin2}) in the cavity $C_j$ satisfies the
orthogonality condition of Eq.~(\ref{ortog}), the state
$\ket{\Psi^{(2)}}$ thus represents entangled orthogonal 2GBSs in two
separate cavities.

For the limit values $p_1,p_2=0,1$, the entangled 2GBSs of
Eq.~(\ref{entbin2}) are reduced to entangled number states having
zero or two photon of the form
{\setlength\arraycolsep{2pt}\begin{eqnarray}
\ket{\Psi^{(2)}}_{p_1=p_2=1}&=&\mathcal{N}\big[\ket{2_10_2}+\eta e^{2i(\phi_2-\phi_1)}\ket{0_12_2}\big],\nonumber\\
\ket{\Psi^{(2)}}_{p_1=1,p_2=0}&=&\mathcal{N}\big[e^{2i(\phi_1+\phi_2)}\ket{2_12_2}+\eta
\ket{0_10_2}\big].\label{ent02}
\end{eqnarray}}This property will be used later on.

\section{Electric field correlations\label{elcorr}}
In order to evidence the non-local properties of the entangled
2GBSs, in this section we examine the electric field correlations
between the two cavities. The quantized electric field inside each
single-mode cavity $C_j$ ($j=1,2$) of frequency $\omega$ and volume
$V$, can be written, at the time $t_j=0$ and in the center of the
cavity, as $\hat{\textbf{E}}_j(z_j)=\textbf{e}_j\hat{E}_j$ where
\cite{vid}
\begin{equation}
\hat{E}_j(z_j)=\sqrt{4\pi\hbar\omega/V}(a_j+a_j^\dag).\label{elec}
\end{equation}
In the following, in order to make a comparison between the
two-photon and the one-photon cases, we first briefly review the
correlations obtained for entangled 1GBSs \cite{lof} and
successively give the results for entangled 2GBSs.

\subsection{Electric field correlations for entangled 1GBSs}
In this case, the two cavities are in the entangled orthogonal 1GBSs
\cite{lof}
\begin{eqnarray}
\ket{\Psi^{(1)}}&=&\mathcal{N}\big[\ket{1,p_1,\phi_1}_1\ket{1,1-p_2,\pi+\phi_2}_2\nonumber\\
&+&\eta\ket{1,1-p_1,\pi+\phi_1}_1\ket{1,p_2,\phi_2}_2\big],\label{entbin1}
\end{eqnarray}
where $\eta$ and ${\cal N}$ are the same of Eq.~(\ref{entbin2}) and
$\ket{1,p_j,\phi_j}_j$ indicates the 1GBS in $C_j$ ($j=1,2$), as
obtained by Eq.~(\ref{bin}). The expectation value
$\bra{\Psi^{(1)}}\hat{E}_j\ket{\Psi^{(1)}}\equiv\ave{\hat{E}_j}$ of
the electric field in $C_j$ is
\begin{eqnarray}
\ave{\hat{E}_j}=4(-1)^{j-1}\sqrt{\frac{\pi\hbar\omega
p_j(1-p_j)}{V}}\frac{1-|\eta|^2}{1+|\eta|^2}\cos\phi_j.\label{AverE}
\end{eqnarray}
$\ave{\hat{E}_j}$ vanishes when $p_1,p_2=0,1$, that is when the
entanglement is between the number states $\ket{0},\ket{1}$, or when
$\modul{\eta}=1$, that is for maximally entangled states.

A quantitative indication of the correlations of electric field
between the cavities is given by the covariance
$\mathcal{C}(\hat{E}_1,\hat{E}_2)=\ave{\hat{E}_1\hat{E}_2}-\ave{\hat{E}_1}\ave{\hat{E}_2}$
that in this case is
\begin{eqnarray}
\mathcal{C}(\hat{E}_1,\hat{E}_2)&=&
\frac{8\pi\hbar\omega}{V}\bigg\{\frac{\eta}{1+|\eta|^2}\big[f(p_1,p_2)\cos\phi_1\cos\phi_2\nonumber\\
&+&\sin\phi_1\sin\phi_2\big]-\left[1-\frac{(1-|\eta|^2)^2}{(1+|\eta|^2)^2}\right]\nonumber\\
&&\times h(p_1,p_2)\cos\phi_1\cos\phi_2\bigg\},\label{Ecov1}
\end{eqnarray}
where
\begin{eqnarray}
f(p_1,p_2)&\equiv&(1-2p_1)(1-2p_2),\nonumber\\
h(p_1,p_2)&\equiv&2\sqrt{p_1p_2(1-p_1)(1-p_2)}.\label{fh}
\end{eqnarray}
${\cal C}(\hat{E}_1,\hat{E}_2)$ is in general different from zero,
and it vanishes when $\eta=0,\pm\infty$, i.e. when the entangled
state $\ket{\Psi^{(1)}}$ of Eq.~(\ref{entbin1}) is reduced to a
product of two uncorrelated 1GBSs. In particular, when $\eta=\pm1$
(maximal entanglement) and $p_1=p_2=1/2$, it is
$\mathcal{C}(\hat{E}_1,\hat{E}_2)=-(4\pi\hbar\omega/V)\cos(\phi_1\pm\phi_2)$.
In this case, if $\phi_1\pm\phi_2=\pi/2$ the covariance vanishes,
while, if $\phi_1\pm\phi_2=0,\pi$, it takes the maximum absolute
value $4\pi\hbar\omega/V$. This shows that the electric fields in
two separate cavities prepared in entangled 1GBSs are correlated and
non-zero.

\subsection{Electric field correlations for entangled 2GBSs}
Using Eqs.~(\ref{elec}) and (\ref{bin}) for $N=2$, the expectation
value of the electric field for the 2GBS $\ket{2,p_j,\phi_j}_j$
($j=1,2$) is
\begin{equation}
{}_j\bra{2,p_j,\phi_j}\hat{E}_j\ket{2,p_j,\phi_j}_j=\sqrt{\frac{2\pi\hbar\omega
p_j(1-p_j)}{V}}\tilde{f}(p_j)\cos\phi_j,\nonumber
\end{equation}
where $\tilde{f}(p_j)\equiv4(1-p_j+\sqrt{2}p_j)$, and it is in
general different from zero, as expected. The two cavities are now
prepared in the entangled 2GBSs of Eq.~(\ref{entbin2}). The mean
electric field
$\bra{\Psi^{(2)}}\hat{E}_j\ket{\Psi^{(2)}}\equiv\ave{\hat{E}_j}$ in
$C_j$ for the state $\ket{\Psi^{(2)}}$ is now given by
{\setlength\arraycolsep{2pt}\begin{eqnarray}
\ave{\hat{E}_j}&=&-\sqrt{\frac{2\pi\hbar\omega
p_j(1-p_j)}{V}}\bigg[\frac{\tilde{f}(j-1+(3-2j)p_j)}{1+|\eta|^{4}}\nonumber\\
&-&\frac{|\eta|^2\tilde{f}(2-j-(3-2j)p_j)}{1+|\eta|^2}\bigg](-1)^j\cos\phi_j.\label{AverE2}
\end{eqnarray}}This mean electric $\ave{\hat{E}_j}$ field in the cavity $C_j$
is in general different from zero when $p_j\neq0,1$ and
$\phi\neq\pm\pi/2$. However, if the entanglement is maximum
($|\eta|=1$), it vanishes only if also $p_j=1/2$. This behavior is
different from the one of the 1GBSs case, where $\ave{\hat{E}_j}$ is
always zero if the entanglement is maximum. In particular, if the
entangled 2GBSs are reduced to the entangled number states of
Eq.~(\ref{ent02}) ($p_j=0,1$), the mean electric field is zero in
each cavity $\ave{\hat{E}_j}$ is zero independently on the value of
$|\eta|$, as expected.

The covariance $\mathcal{C}(E_1;E_2)$ of the electric fields for
entangled 2GBSs is
{\setlength\arraycolsep{2pt}\begin{eqnarray}
\mathcal{C}(E_1;E_2)&=&\frac{\pi\hbar\omega h(p_1,p_2)}{V}
\bigg\{\bigg[\frac{F(p_1)F(1-p_2)}{(1+|\eta|^2)^{2}}\nonumber\\
&-&\frac{\widetilde{F}(p_1,p_2)+|\eta|^2\widetilde{F}(p_2,p_1)}{1+|\eta|^2}\bigg]\cos\phi_1\cos\phi_2\nonumber\\
&-&\frac{8|\eta|(3-2\sqrt{2})}{1+|\eta|^2}\overline{F}(p_1,p_2,\phi_1,\phi_2)\bigg\},\label{Cov2}
\end{eqnarray}}
where
{\setlength\arraycolsep{2pt}\begin{eqnarray}
F(p_j)&=&\tilde{f}(p_j)-|\eta|^2\tilde{f}(1-p_j),\nonumber\\
\widetilde{F}(p_1,p_2)&=&\tilde{f}(p_1)\tilde{f}(1-p_2),\nonumber\\
\overline{F}(p_1,p_2,\phi_1,\phi_2)&=&f(p_1,p_2)\cos\phi_1\cos\phi_2+\sin\phi_1\sin\phi_2,\nonumber
\end{eqnarray}}and $f(p_1,p_2),h(p_1,p_2)$ are defined by Eq.~(\ref{fh}).
The covariance vanishes when $p_1,p_2=0,1$. Thus, the electric
fields in the two cavities are always uncorrelated for entangled
zero and two-photon number states, as given by Eq.~(\ref{ent02}).
The covariance $\mathcal{C}(E_1;E_2)$ also vanishes when
$\eta=0,\pm\infty$, that is when there is no entanglement between
the 2GBSs of the two cavities. In particular, for $|\eta|=1$,
$p_1=p_2=1/2$ and $\phi_1=\phi_2=\phi$, the covariance of
Eq.~(\ref{Cov2}) becomes
$\mathcal{C}(E_1;E_2)=-2\pi\hbar\omega(3+2\sqrt{2}\cos2\phi)/V$,
that is different from zero, independently from the value of the
mean phase $\phi$ appearing in the entangled 2GBSs of
Eq.~(\ref{entbin2}). This behavior is different from the one of
entangled 1GBSs having the same values of the parameters, where the
covariance becomes constant and equal to $-4\pi\hbar\omega/V$ when
$\eta=-1$, while it is $-(4\pi\hbar\omega/V)\cos2\phi$ when
$\eta=+1$. Therefore, preparing entangled 1GBSs or entangled 2GBSs
with given values of the characteristic parameters $p,\phi$
\cite{lof,lof1}, we can obtain a different behavior of the
covariances in the two cases. Thus, we have shown that the electric
fields of two cavities, prepared in entangled 2GBSs, are correlated.

\section{Bell's inequality violation\label{bellviol}}
In the previous section we have found that the electric fields of
two cavities, prepared in entangled 2GBSs, are correlated. In this
section, instead, we shall analyze the quantum nonlocal correlations
for entangled 2GBSs by using Bell's inequality in the
Clauser-Horne-Shimony-Holt (CHSH) form \cite{clau,mand}. To this
purpose, we introduce a measurable dichotomic operator expressed in
terms of the cavity field states.

The two orthogonal 2GBSs $\ket{2,p,\phi},\ket{2,1-p,\pi+\phi}$,
constitute the basis vectors of a 2-dimensional subspace,
$\mathcal{B}=\{\ket{2,p,\phi}\equiv\ket{+},\ket{2,1-p,\pi+\phi}\equiv\ket{-}\}$,
of the 3-dimensional Hilbert space \cite{lof2}. We therefore
construct, for each single-mode cavity, a dichotomic operator
$\hat{F}$, acting within the 2-dimensional subspace spanned by the
basis $\mathcal{B}$. Defining a 3-dimensional vector
$\vec{F}\equiv(F_x,F_y,F_z)$, we construct the operator $\hat{F}$
\begin{equation}
\hat{F}=\vec{F}\cdot\vec{\sigma}=\left(\begin{array}{cc}F_z&F_x-iF_y\\F_x+iF_y&-F_z\end{array}\right)
\equiv\left(\begin{array}{cc}F_{11}&F_{12}\\F_{12}^\ast&-F_{11}\end{array}\right),\nonumber
\end{equation}
where $\vec{\sigma}=(\sigma_1,\sigma_2,\sigma_3)$ are the Pauli
matrices acting on the $\mathcal{B}$ subspace while the parameters
$F_z,F_{12}$ will be shown to be linked to the coefficients of the
linear superposition of the basis states. In order that this
dichotomic operator $\hat{F}$ has eigenvalues $\pm1$, it must be
\begin{eqnarray}
|\vec{F}|^2&=&F_x^2+F_y^2+F_z^2=|F_{12}|^2+F_z^2=1\Rightarrow\nonumber\\
|F_{12}|&=&\sqrt{1-F_z^2},\quad
F_{12}=|F_{12}|e^{i\vartheta}.\label{F12}
\end{eqnarray}
The expression of the operator $\hat{F}$ in terms of the basis
vectors is then
\begin{eqnarray}
\hat{F}&=&F_z(\ket{+}\bra{+}-\ket{-}\bra{-})\nonumber\\
&+&\sqrt{1-F_z^2}(e^{i\vartheta}\ket{+}\bra{-}+e^{-i\vartheta}\ket{-}\bra{+}),\label{F}
\end{eqnarray}
and its eigenvectors are given by
\begin{eqnarray}
\ket{\widetilde{+}}&=&[\sqrt{1+F_z}\ket{+}+\sqrt{1-F_z}e^{-i\vartheta}\ket{-}]/\sqrt{2},\nonumber\\
\ket{\widetilde{-}}&=&[-\sqrt{1-F_z}e^{i\vartheta}\ket{+}+\sqrt{1+F_z}\ket{-}]/\sqrt{2}.\label{eigenvectors}
\end{eqnarray}
We shall show that this operator satisfies a CHSH-Bell inequality
violations for entangled 2GBSs.

Let us consider the entangled 2GBSs $\ket{\Psi^{(2)}}$ given in
Eq.~(\ref{entbin2}) and take $p_1=p_2=p$ and $\phi_1=\phi_2=\phi$.
This choice shall simplify the expressions, making the results more
easily readable. The operator of Eq.~(\ref{F}) in the cavity $C_j$
($j=1,2$) is indicated as $\hat{F}^{(j)}(\vartheta_j)$, with $F_z$
being equal in the two cavities. The quantum correlation of the
operator $\hat{F}$ in the two cavities is defined as
$\ave{\hat{F}^{(1)}(\vartheta_1)\hat{F}^{(2)}(\vartheta_2)}=\bra{\Psi^{(2)}}\hat{F}^{(1)}(\vartheta_1)\hat{F}^{(2)}(\vartheta_2)\ket{\Psi^{(2)}}$,
and it results to be
\begin{equation}
\ave{\hat{F}^{(1)}(\vartheta_1)\hat{F}^{(2)}(\vartheta_2)}=\frac{2\eta(1-F_z^2)}{1+|\eta|^2}\cos(\vartheta_1-\vartheta_2)-F_z^2.\label{corrF}
\end{equation}
In terms of these correlations it is possible to construct the
CHSH-Bell inequality as \cite{clau,mand}
\begin{eqnarray}
S_B=\modul{\ave{\hat{F}^{(1)}(\vartheta_1)\hat{F}^{(2)}(\vartheta_2)}
-\ave{\hat{F}^{(1)}(\vartheta_1)\hat{F}^{(2)}(\vartheta'_2)}}\nonumber\\
+\modul{\ave{\hat{F}^{(1)}(\vartheta'_1)\hat{F}^{(2)}(\vartheta_2)}
+\ave{\hat{F}^{(1)}(\vartheta'_1)\hat{F}^{(2)}(\vartheta'_2)}}\leq2.\label{BellF}
\end{eqnarray}
Thus, $S_B$ is formed by correlations of the kind (\ref{corrF}),
having different phase angles $\vartheta_j,\vartheta'_j$ but the
same dependence on $F_z$. At this point, we look for opportune
values of the parameters $\eta,F_z,\vartheta_j,\vartheta'_j$ of
Eq.~(\ref{BellF}) such that $S_B>2$ and thus the Bell's inequality
violation occurs. Setting the partial derivative relating to $F_z$
of the correlation functions appearing in Eq.~(\ref{BellF}) equal to
zero, we readily obtain
\begin{equation}
\frac{\partial S_B}{\partial
F_z}=F_zf(\eta,\vartheta_1,\vartheta_2,\vartheta'_1,\vartheta'_2)=0\Rightarrow
F_z=0,\label{derivative}
\end{equation}
where the function
$f(\eta,\vartheta_1,\vartheta_2,\vartheta'_1,\vartheta'_2)$ is never
zero. It is possible to see that $F_z=0$ corresponds to a maximum of
$S_B$. For this value of $F_z$, the CHSH-Bell inequality of
Eq.~(\ref{BellF}) becomes
\begin{eqnarray}
S_B&=&G^{(E)}[|\cos(\vartheta_1-\vartheta_2)-\cos(\vartheta_1-\vartheta'_2)|\nonumber\\
&+&|\cos(\vartheta'_1-\vartheta_2)-\cos(\vartheta'_1-\vartheta'_2)|]\leq2,
\label{SB}
\end{eqnarray}
where $G^{(E)}=2|\eta|/(1+|\eta|^2)$ is the degree of entanglement
\cite{abour}, invariant with respect to the substitution
$\modul{\eta}\rightarrow1/\modul{\eta}$, equal to zero for
$\modul{\eta}=0,+\infty$ (uncorrelated states) and equal to one for
$\modul{\eta}=1$ (maximally entangled states). Choosing appropriate
values of the angles $\vartheta_j,\vartheta'_j$, the $S_B$ function
of Eq.~(\ref{SB}) can be shown to take values greater than two, so
that the CHSH-Bell inequality is violated.

\begin{figure}
\includegraphics[width=0.46\textwidth]{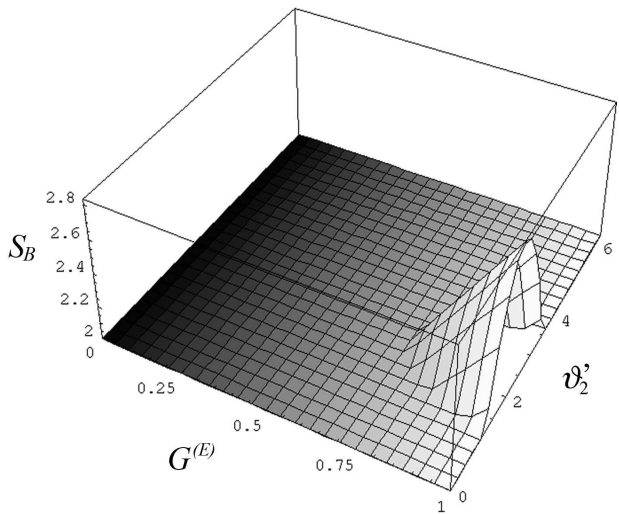}
\caption{\label{fig1}Plot of $S_B$ as a function of the degree of
entanglement $G^{(E)}$ and the operator angle $\vartheta'_2$. The
other operator angles are fixed at $\vartheta_1=0$,
$\vartheta_2=\pi/4$ and $\vartheta'_1=\pi/2$. The $S_B$ axis starts
from the maximum classical limit $S_B=2$, while $0\leq G^{(E)}\leq1$
and $0\leq\vartheta'_2\leq2\pi$. A Bell's inequality violation
occurs when $S_B>2$.}
\end{figure}
For example, after choosing $\vartheta_1=0$, $\vartheta_2=\pi/4$,
$\vartheta'_1=\pi/2$, we plot $S_B=S_B(G^{(E)},\vartheta'_2)$ in
Fig.~\ref{fig1}, which shows that for some values of $\vartheta'_2$
and $G^{(E)}$, we have $S_B>2$. The CHSH-Bell inequality is violated
also when the degree of entanglement $G^{(E)}$ is not maximum. In
particular, when $\vartheta'_2=3\pi/4$, from Eq.~(\ref{SB}) we
obtain
\begin{equation}
S_B=2\sqrt{2}G^{(E)}>2\Rightarrow1/\sqrt{2}\approx0.707<G^{(E)}\leq1.
\end{equation}
When the degree of entanglement is maximum ($G^{(E)}=1$), the
maximum value of $S_B$ ($S_B^{max}=2\sqrt{2}\approx2.828$) is
obtained. This value represents the maximal possible violation of
the CHSH-Bell inequality \cite{niels}.

It is important to note that the choice of the operator parameters
$F_z,\vartheta$ in each cavity determine the eigenvectors of the
operator $\hat{F}$ of Eq.~(\ref{F}), as readily seen from
Eq.~(\ref{eigenvectors}). These eigenvectors represent states of the
cavity field expressed as superpositions of two orthogonal 2GBSs. It
is possible to show that these field states can be in principle
measured by probe two-level atoms that ``read'' the cavity field
\cite{lof3}. The possibility of measuring the eigenvectors of the
cavity operator $\hat{F}$, corresponding to the measurement of its
eigenvalues $\pm1$, opens thus the way to an experimental Bell test
for entangled 2GBSs in two separate cavities. The correlations
$\ave{\hat{F}^{(1)}(\vartheta_1)\hat{F}^{(2)}(\vartheta_2)}$ for the
desired values of the angles $\vartheta_1,\vartheta_2$ can be
obtained by statistical averages on the ensemble of the
measurements. The CHSH-Bell inequality of Eq.~(\ref{BellF}) can be
thus finally tested.

\section{Conclusion\label{concl}}
In this paper we have analyzed the non-local properties of entangled
two-photon generalized binomial states (2GBSs) in two spatially
separate cavities. In particular, we have investigated the
expectation values and the correlations of the electric field for
these entangled states. We have also compared these results with the
ones for entangled 1GBSs, emphasizing the different behavior of the
quantum correlations in the two cases.

We have constructed a Bell's inequality by using an appropriate
dichotomic cavity field operator that is in principle measurable by
probe two-level atoms. We have then shown that the CHSH-Bell
inequality applied to entangled 2GBSs can be violated for a wide
range of the degree of entanglement (Sec.~\ref{bellviol}). We point
out here that the atomic state detector efficiency $\alpha$ plays an
important role in the experimental realization of a Bell test. Here
we have supposed an ideal efficiency ($\alpha=1$) (see
Sec.~\ref{bellviol}), but if we include the detectors efficiencies
in the correlation functions, the CHSH-Bell inequality would not be
violated for values of $\alpha$ less than a threshold value
$\alpha_t\approx0.8284$ for maximally entangled states
\cite{clau,massar}. However, this problem can be overcome by the
so-called ``fair sampling'' hypothesis, where the sub-ensemble of
detected events (detected probe atoms) represents the whole
ensemble. Thus, the results rely only on the detected events, but
the detection loophole remains ``open'' \cite{moeh,massar}. Only for
detector efficiencies greater than $\alpha_t$ the detection loophole
can be closed. It is worth to note that it could be very difficult
to realize experimental loophole free Bell tests, because we would
need simultaneous and perfect efficiency measurements of eventual
probe atoms. Anyway, recent laboratory developments open promising
perspectives for a better and easy control of a well-defined atom
numbers sequence \cite{maioli} and for a high atomic detection
efficiency in microwave CQED experiments \cite{auff}. The
realization of the Bell test proposed here for entangled 2GBSs would
give a direct demonstration of non-local behavior for entangled
cavity fields with a photon number $N>1$.

\end{document}